\documentclass[acmsmall]{acmart}
\AtBeginDocument{%
  }

\renewcommand\footnotetextcopyrightpermission[1]{}
\usepackage[table]{xcolor}% http://ctan.org/pkg/xcolor
\usepackage{tabularx}
\usepackage[dvipsnames]{xcolor}
\usepackage{makecell}
\usepackage[acronym,nomain,nonumberlist]{glossaries}
\usepackage{hyperref}
\usepackage[font=small,labelfont=bf,skip=2pt]{caption}
\makeglossaries

\newacronym{ai}{AI}{artificial intelligence}
\newacronym{bsp}{BSP}{bulk synchronous parallel}
\newacronym{mpi}{MPI}{Message Passing Interface}
\newacronym{hpc}{HPC}{high performance computing}
\newacronym{aws}{AWS}{Amazon Web Services}
\newacronym{gke}{GKE}{Google Kubernetes Engine}
\newacronym{aks}{AKS}{Azure Kubernetes Service}
\newacronym{eks}{EKS}{Elastic Kubernetes Service}
\newacronym{ml}{ML}{machine learning}
\newacronym{rdma}{RDMA}{Remote Direct Memory Access}
\newacronym{vm}{VM}{virtual machine}
\newacronym{fom}{FOM}{figure of merit}
\newacronym{efa}{EFA}{Elastic Fabric Adapter}
\newacronym{ec2}{EC2}{Elastic Compute Cloud}
\newacronym{ucx}{UCX}{Unified Communication X}
\newacronym{cni}{CNI}{container networking interface}
\newacronym{api}{API}{application programming interface}
\newacronym{cagr}{CAGR}{compound annual growth rate}

\begin{document}

%%
%% The "title" command has an optional parameter,
%% allowing the author to define a "short title" to be used in page headers.
\title{Possible Futures for Cloud Cost Models}

%%
%% The "author" command and its associated commands are used to define
%% the authors and their affiliations.
%% Of note is the shared affiliation of the first two authors, and the
%% "authornote" and "authornotemark" commands
%% used to denote shared contribution to the research.
\author{Vanessa Sochat}
%\authornote{corresponding author}
\email{sochat1@llnl.gov}
\orcid{0000-0002-4387-3819}
\affiliation{%
  \institution{Lawrence Livermore National Laboratory}
  \city{Livermore}
  \state{CA}
  \country{USA}
}

\author{Daniel Milroy}
\email{milroy1@llnl.gov}
\orcid{0000-0001-6500-3227}
\affiliation{%
  \institution{Lawrence Livermore National Laboratory}
  \city{Livermore}
  \state{CA}
  \country{USA}
}

%%
%% By default, the full list of authors will be used in the page
%% headers. Often, this list is too long, and will overlap
%% other information printed in the page headers. This command allows
%% the author to define a more concise list
%% of authors' names for this purpose.
\renewcommand{\shortauthors}{Sochat and Milroy.}

%%
%% The abstract is a short summary of the work to be presented in the
%% article.
\begin{abstract}
Cloud is now the leading software and computing hardware innovator, and is changing the landscape of compute to one that is optimized for artificial intelligence and machine learning (AI/ML). Computing innovation was initially driven to meet the needs of scientific computing. As industry and consumer usage of computing proliferated, there was a shift to satisfy a multipolar customer base. Demand for AI/ML now dominates modern computing and innovation has centralized on cloud. As a result, cost and resource models designed to serve AI/ML use cases are not currently well suited for science. If resource contention resulting from a unipole consumer makes access to contended resources harder for scientific users, a likely future is running scientific workloads where they were not intended. In this article, we discuss the past, current, and possible futures of cloud cost models for the continued support of discovery and science.
% Cloud is now the leading software and computing hardware innovator, and is changing the landscape of compute to one that is optimized for artificial intelligence and machine learning (AI/ML). In the past, scientific computing was a primary, single pole that drove innovation throughout all of computing.  As industry and consumer-usage of computing proliferated, there first was a shift to a multi-polar customer base that broadly encompassed science, consumer, and commercial entities. As the demand of the AI/ML market takes predominant focus, innovation has shifted back to a single pole driven by industry. Cost and resource models that are geared toward these AI/ML use cases are not currently well suited for science.  If resource contention makes access to contended resources harder for scientific users, a likely future is running scientific workloads where they were not intended. In this article, we discuss the past, current, and possible futures of cloud cost models for the continued support of discovery and science.
\end{abstract}

% We started with a polar world for customer base - innovation was directed for that pole. We had a shift to have a tri-polar world with three customer bases. We now are shifting back to a single polar world, shifting totally toward AI/ML consumer.

% NOTE FROM V: I think we need to do a better job of saying WHY the current models aren't good for science.

%%
%% The code below is generated by the tool at http://dl.acm.org/ccs.cfm.
%% Please copy and paste the code instead of the example below.
%%


%%
%% Keywords. The author(s) should pick words that accurately describe
%% the work being presented. Separate the keywords with commas.

%\received{20 June 2025}
%\received[revised]{12 July 2025}
%\received[accepted]{5 September 2025}

%%
%% This command processes the author and affiliation and title
%% information and builds the first part of the formatted document.
\maketitle

\section{Introduction}

Cloud computing is a service model that promises access to computing resources (compute, storage, networking, and services) for individuals and companies alike. The benefits to the consumer include not having to manage the underlying resources, and the ability to request resources when needed. Ideally, cloud resource provisioning is provided at a price that is more affordable, sustainable, and resource efficient than having to purchase, power, and maintain resources on premises.  While services offered and rates have changed since its inception in the early 2000s, the cost models for cloud computing are limited to on-demand, reservation, and committed use. With the marked increase for highly contended resources such as GPUs comes the need for more sophisticated models of resource scheduling and cost to effectively satisfy the demand. Optimizing access to compute resources is increasingly important due to proposed cuts to scientific budgets ~\cite{Vought2025-hd}. 

% "changed with the same degree of fluctuation" we probably should try to quanity this.

% Note from Dan about "ideally this provisioning"
% purchase, power, and maintain. It may be good to mention carbon footprint, as clouds tout their environmental efficiency and provide options to prioritize lower footprint.

%% Isn't some of it still federally funded?
Infrastructure for research is often federally funded \cite{Committee-on-Innovations-in-Computing-and-Communications:-Lessons-from-History1999-un}. Much of innovation in computing has relied on academic, national labs, and private research institutions receiving funding from the government and private companies. The incentive for private companies to invest in the expertise of the academic sector has been provided by technology transfer and the eventual commercialization of the research. However, as cloud vendors become the predominant leaders in the computing economy, with a 20\% \gls*{cagr} that is expected to reach over \$1.28T USD of revenue by 2028 \cite{gartner-2024}, cloud vendors do not need to collaborate to innovate. They can hire the talent they need, and purchase or develop resources in-house. The \gls*{hpc} market is smaller but growing, with total revenue of \$60B USB in 2024, and projected growth to \$100B by 2028 \cite{hyperion-hpc-ai}. Despite its growth, the \gls*{hpc} community faces greater challenges to bring in enough funds to sustain itself. The long time scale of research returns, the risk due to the uncertain outcome, and the disconnection from the economy raises questions about the value of research.  The disconnect between this value and economic benefit of its primary customer base makes it harder for \gls{hpc} to obtain funding.

% The challenges are due in part to the uncertain returns of its scientific customer base and the long time scales that % The current political landscape adds further uncertainty about this future, with recent action to cut models of scientific funding that have been relied upon for decades \cite{Jahnke2015-so}.

The return on investment (ROI) that results from science is rare and often takes decades to manifest. The Global Positioning System (GPS) is one example that grew out of Einstein's Theories of Special and General Relativity (1905 and 1915). GPS alone has led to a market estimated at \$150 billion annually, and a total economic impact in the trillions of dollars. Transitors follow a similar pattern, with ideas originating in quantum mechanics work from the 1920's that is now a \$20 trillion market almost a century later. Other scientific discovers that have significant ROI include RSA encryption (global e-commerce that uses it alone is over \$25 trillion), genetic engineering (the biotechnology industry is valued at over \$1 trillion), and laser technologies (trillions). The benefits of science are often not seen for decades or even a century later. An industry-oriented, profit-driven company will not invest in a payout at that long a timescale.

% See bottom of here: https://gist.github.com/vsoch/87a347bad04556172a4aae82db885e16 for references - I didn't add any due to space but maybe we can finagle them around.

Given the demand for resources that power \gls*{ai} \cite{ClearMLUnknown-gc} and a relatively smaller \gls*{hpc} market, the scientific community may come to rely on cloud vendors for scientific computing. However, it is uncertain if scientific demand can be met for smaller, academic groups~\cite{access-initiative} given the overall demand. Scientific access to contended resources would be further limited if there is competition for the underlying resources in the supply chain (Section \ref{section:supply-chain}). It may not be possible to purchase \gls*{hpc} clusters at the capacity needed by certain scientific workloads. Scientific workloads must best utilize all available resources, including current and future cloud offerings. The use of cloud resources is subject to current cloud cost models, which do not empower research groups to obtain resources for time periods comparable to batch jobs. On-demand models no longer function well under resource contention \cite{sochat2025usabilityevaluationcloudhpc}, and reservation models that require huge investment are not tenable for small research groups. Thus, the current economic and computational landscape is not supportive of the future of science.  % While efforts to approach this challenge from the national level are admirable and should not be abandoned \cite{Deelman2025-ox}, in the case that federal funding cannot be relied upon, 
In the case that public or federal funding cannot be relied upon, direct collaboration and discussion with cloud vendors are a possibility, and arguably a more direct path. Discussion of past and present cost models and a clear definition of what exists and is currently missing is needed. In this paper, we review the early definition of cloud computing and cost models that support it, discuss why the current state does not work with these models, and offer ideas for possible futures. 

\subsection{Emergence of Infrastructure As a Service}
While the concept of cloud computing predates the previous 25 years \cite{Varghese2019-nc}, in 1999 Amazon wanted to transform their book-selling service into something more, and in 2002 they launched their first public cloud \gls*{aws}~\cite{UnknownUnknown-rc}. The sales pitch for cloud computing was to ``instantly spin up hundreds or thousands of servers in minutes and deliver results faster,''  suggesting immediate access to large numbers of resources. This branding has persisted for more than two decades, and many still present the cloud as having infinite resources. In practice, this has been shown to not be the case for \gls{hpc} workloads \cite{Lange_undated-qa,sochat2025usabilityevaluationcloudhpc}.
% , and arguably there is little incentive to dispel the erroneous message because the illusion of infinite resources supports profitable branding and consumer image. 
% The nuanced detail is that this statement used to be true, but no longer is given the change in the kinds of compute needed, and the amount, a point we will further unwrap later in this writing.

In 2006 Amazon released the first commercial cloud, the \gls*{ec2}, that offered access to one or more \gls*{vm} instances and storage. Companies could pay \gls*{aws} to host their services, which customers could access via the internet and mobile devices \cite{Varghese2019-nc}. Google followed in 2008 with Google App Engine, and Microsoft a few months later with Windows Azure. This model was referred to as IaaS ``Infrastructure as a Service'' and broadly promised that users could get the resources they needed in the quantities required whenever they needed them.  Cloud and \gls{hpc} initially served separate customer bases, but as cloud revenue grew, it began to expand into \gls{hpc}. Biosciences were the first scientific community to create workflow tools oriented to use \gls*{vm}s offered by public cloud vendors \cite{Koster2012-hv}. 

% the idea of scientific clouds was not too far out, with the first scientific cloud ``Nimbus'' \cite{Keahey2008-gn} being prototyped in 2008, and shortly followed by workflow tools oriented to use \gls*{vm}s offered by public cloud vendors \cite{Koster2012-hv}.

% Professor Dan Reed talks about \gls*{ec2} as early as 2007, and alludes to the perception of infinite resources, describing running image using ``as many or as few systems as you desire.'' \cite{reed-ruminations}

%\begin{quote}
%Amazon EC2 presents a true virtual computing environment, allowing you to use web service interfaces to request machines for use, load them with your custom application environment, manage your network's access permissions, and run your image using ``as many or few systems as you desire'' \cite{reed-ruminations}. 
%\end{quote}

% With annual projections for cloud services growth of between 15-20\% \cite{gartner-2024}, tech companies continue to invest in cloud offerings \cite{Russell2023-zk}. Arguably, in the early days of cloud computing, small companies might have benefited most from these services, as they could not afford their own data centers. 

% I'm not sure where to best put this.

\subsection{Models Needed for Science}
Science cost models support workload patterns based on ephemeral funding, sporadic work, and research outcomes. Standard business models based on commercial traffic call for persistent services and long-term discounts for purchasing resources. In contrast, scientific runs are typically short and infrequent. A scientist might need a cluster with specialized, high-precision hardware a few times a month to run a large simulation. In the case of larger scientific simulations, cost models based on saving money with preemptible instances can be risky, as the failure of one instance can lead to failure of the entire simulation because of the rigidity of \gls{mpi}.

% Question - if scientific funding is based on grants or planning, why don't the reservation models work? Just wrong instance types or scale?

Profits from commercial entities can typically be reinvested and used to continue spend on cloud resources. This is not the case for scientific projects, which are backed by one-time use, soft money like grants. 
% A promise of ROI also cannot be used as a carrot. Science is measured in discovery, not profit. The ROI of a fundamental scientific discovery can take years, and even decades, to lead to economic or social ROI, often unbeknownst to the discoverer. 
Scientists who receive funds from scientific grants may use them toward cloud, but cannot always repeat the purchase or dictate direction for broader institutional purchases. When a cloud vendor supports the computing needs of a research group through credits, the vendor can expect the support to transition to a profitable business relationship. Long-term profitability may not be feasible if the research group has limited influence over its parent institution's procurements.

% Other ideas: "Cost comparison ability" E.g., "What if I used GPUs instead of CPUs? What if I used instance X instead of Y?"
% Informed spot: understanding the probability of interruption to be able to figure out when spot makes sense.
% Data lifecycle or storage cost modeling

\section{Current Challenges}

NIST defined five essential characteristics of cloud computing in September 2011, including \emph{on-demand self-service}, \emph{broad network access}, \emph{resource pooling}, \emph{rapid elasticity}, and \emph{measured service} \cite{MellUnknown-pj}. % They also defined two service models, software as a service (SaaS) and platform as a service (PaaS), that describe using and deploying applications on cloud infrastructure. Arguably, cloud has had greater utility for consumers by way of the second model, as it is more typical for a company to deploy their own applications. 
We evaluate the current landscape in context of these definitions and the needs of science.

\label{section:supply-chain}
\subsection{Current cloud cost models are challenged by chip shortages}

% https://www.eetimes.com/global-foundry-market-navigates-milder-dip-in-2025/

\gls*{ai}/\gls*{ml} is projected to grow from \$123.16 billion USD (2024) to \$311.58 billion by 2029, a \gls*{cagr} of 20.4\% \cite{Rayner2024-to}. Cloud hyperscalers are the primary providers of AI/ML infrastructure and can translate the demand for AI/ML into large-scale chip purchases. Fabricating semiconductors has a high barrier to entry and is thus dominated by a small number of companies, including the Taiwan Semiconductor Manufacturing (TSMC) with 67.6\% of market share, Samsung (8.1\%), SMIC (5.5\%), UMC (4.7\%), among others \cite{News-Desk2025-rh}. There are high barriers to entry for chip manufacturing, with the cost of a fabrication utility ``fab'' ranging from 5 to upward of 20 billion dollars per unit \cite{Sevitz2024-yw}. The small number of suppliers combined with exploding demand, resource shortages \cite{Unknown2024-rk}, and issues in production or policy can lead to shortages, such as those experienced during the COVID-19 pandemic. This shortage led to increased prices and lower availability of cars, compute and other electronic devices. High-end chips continue to be contended resources \cite{Werner2024-qb}. % Reliable manufacturing and an adequate labor force may not be established for years, and the shortage is exacerbated by geopolitical events . 
% Demand for chips is larger than can be produced, and challenged by weaknesses in the supply chain \cite{Unknown2024-rk}. 
Increased tariffs could further increase chip cost, placing more supply-side responsibility on local manufacturing that cannot yet meet demand. The current state is a limited supply of chips that results from a small number of manufacturers, political barriers, and the limitations of physical resources \cite{Parvini2025-nd}. %It cannot be known how the future will play out. % \cite{Parvini2025-nd}. % The political uncertainty could lead to an increased reliance on commercial cloud.

The on-demand model of resource provisioning does not work without sufficient supply. The most expensive resources, including GPU and some CPU instances, are often unavailable at the scale needed for scientific simulation, an experience that can be observed across clouds \cite{sochat2025usabilityevaluationcloudhpc,Lange_undated-qa}. Customers are often required to first request quota for a resource, which is typically a manual process. We can only speculate on the reasons why. The first possibility is control of resources -- a cloud resource that is in limited supply and highly demanded needs to be intelligently managed. The second might be legal protection for accidental spending. When a customer requests a quota, there is a record of the request and outcome. Finally, the manual process of a quota request might make it slightly harder for malicious automated account creation. The NIST definition of cloud computing requires that provisioning occurs ``without requiring human interaction,'' yet many operations in a cloud account require manual interaction with the vendor. Provisioning resources in many cases requires special connections, networks, large sums of money, or in-person meetings to finalize contracts. Even when a vendor grants a quota request, the actual capacity is not guaranteed to be available and customers may be subject to preferential treatment.

% \label{section:resource-efficiency}
% \subsection{Resource and environmental efficiency}

% purchase, power, and maintain. It may be good to mention carbon footprint, as clouds tout their environmental efficiency and provide options to prioritize lower footprint.

\subsection{Resource Pooling}
While clouds serve multiple consumers typically via a project-oriented model, the details to that multi-tenancy setup are opaque to the user. In an \gls*{hpc} system there is more transparency with fair-share algorithms for requesting work. In cloud, it is likely the case that customer tiers exist based on account reputation and priority of resource access based on spend. If the essence of resource sharing as defined by NIST is to assign and reassign according to demand without unfairly giving priority, it is not clear that commercial models could be true to this point. If we consider customer tiers and inequality of access as a reasonable attribute of a commercial system, then it holds. The ability to get resources can vary based on an account type \cite{sochat2025usabilityevaluationcloudhpc}. % An account funded by credits, in our experience, was much more challenging to get resources for than a fully paid account, despite the credit account existing for much longer, and thus having more time to form a trusted reputation.

\subsection{Rapid Elasticity}
The claim of rapid elasticity encompasses a lot of the same promises of on-demand resources, specifically that, ``to the consumer, the capabilities available for provisioning often appear to be unlimited and can be appropriated in any quantity at any time.'' % This is no longer the current state of the world for reasons previously mentioned (Section \ref{section:supply-chain}). 
The NIST definition ``to scale rapidly outward and inward commensurate with demand'' mandates scaling as a characteristic of cloud. While cluster creation is reasonably quick (e.g., Terraform deployments across clouds deploy \gls*{vm}s in minutes, and the fastest Kubernetes deployments can take 5-6 minutes), problems arise when capacity is not available. In this case a subset of resources might be provisioned, and then a waiting period up to 30-35 minutes is allowed before the operation fails. In a recent performance study \cite{sochat2025usabilityevaluationcloudhpc} the inability to meet the minimum number of required resources incurred a charge of \$4000 while waiting for nodes that were never allocated. Vendors charging for idle resources is not intentional, but results from the deficiency of the cost and allocation models. % Whether the NIST requirement of ``rapidly'' is met is subjective, and up to the assessment of the reader. 
 
% assuming that resources are available. Adding or removing nodes from a cluster can take between 30 seconds and a few minutes, also depending on the cloud and mechanism used. In the case that capacity is not available, 

% for \gls*{aws} native APIs (3-4 minutes), CloudFormation (15-25 minutes),  For example, using native APIs in \gls*{aws}, a cluster on \gls*{ec2} can be provisioned in about 3 minutes. An equivalent cluster, resource-wise, provisioned with the service CloudFormation can take between 15-25 minutes. It is unclear why there is this discrepancy, but it can be guessed that the latter is not as efficient with the methods used for the creation. In comparison, creating a cluster of the same service type (e.g., Kubernetes) 
\subsection{Measured Service}

The ``measured service'' characteristic of the NIST definition of cloud computing mandates transparency of usage and cost in cloud computing. The document states that ``Resource usage can be monitored, controlled, and reported, providing transparency for both the provider and consumer of the utilized service'' \cite{MellUnknown-pj}. In practice, consumers actively using resources do not have awareness about cost or metrics until the next business day or longer. The lack of immediate cost transparency makes it very challenging to make cost-effective planning in the moment, and often burdens the consumer to do research and plan experiments carefully in advance. A user coming from academia may be unfamiliar with this new model because planning for budget is uncommon. Inexperience can lead to unexpected unintended spending that can exceed budgets and discourage further usage.

\section{Past and Current Models}

\subsection{The cloud promise}
``Service Level Agreements'' or SLAs help clarify the promises cloud vendors make to customers about resources. As an example, for \gls*{aws} \gls*{ec2}, the promise is not about the performance of a type of resource, but rather that a certain percentage of an instance type is available in a given region \cite{UnknownUnknown-vf}. The idea of an instance being unavailable does not pertain to being able to get it, but rather failing to meet a minimum capability or quality of service (QoS), which is stated to relate to external connectivity. The QoS has force majeure clauses for events outside of \gls*{aws}' control. 

Google Cloud maintains similar SLAs for their Compute Engine services \cite{UnknownUnknown-zy}, with a notable difference of an explicit declaration of tiers of service, including Premium and Standard. Unlike \gls*{aws}, their SLAs define the concept of ``downtime'' that includes external connectivity, extends to load balancing, and does not cover specific cases for VPN. Periods of small downtime are not covered. SLA violations reimburse customers with credits to reuse on the cloud platform rather than reimbursement in the initial form of payment. % The idea behind this model is likely that the customer has committed the funds to being used on the cloud, and can try again but cannot be refunded back to the initial payment method.

Interestingly, there is no guarantee that the user receives consistent hardware that is posted on a website. This might be called the ``supermarket fish problem,'' where the consumer purchases a generic kind of instance ``white fish'' and then can ultimately get multiple different micro-architectures \cite{sochat2025usabilityevaluationcloudhpc}. These authors have observed this behavior only with respect to different scales of requests, where a single instance might be provisioned with the latest generation of an architecture, and a large set with an older variant. Cloud vendors likely make the best effort to provide consistent types, and variation is a result of rapidly changing updates to data center machines that are generally beneficial and would be challenging to compensate for~\cite{sochat2025usabilityevaluationcloudhpc}.
 % This behavior has been observed directly by these authors for N2 instance types, where the underlying architecture can vary in generation based on the scale needed. % In practice, the authors note that subjectively, it is unlikely to get an allocation or group with different instance types. The issue comes up when the predominant instance architecture changes over time or in different contexts, resulting in changes to a previously run workload or even segfaults. 
%Experiments to look at underlying architectures show relative consistency, and these authors believe that cloud vendors make a best effort to provide consistent types, and variation is a result of rapidly changing updates to data center machines that are overall beneficial and would be challenging to offset.

\subsection{On-demand and pay-as-you-go}
Pay-as-you-go pricing \cite{Unknown2023-kg} is typically paired with on-demand resources, meaning that a consumer requests a count of a resource type, and then pays for a duration of usage. One advantage of pay-as-you-go pricing is knowing that the final cost depends only on exactly what is used. % and little planning is required in advance to provision resources. 
Ideally, the user requests resources when they are needed, and they are immediately provisioned in response. A higher monetary cost comes with this convenience. As an example, at the time of this paper, the \gls*{aws} \emph{p5.48xlarge} instance type can cost between \$55.04  to \$92.46 per hour when allocated on-demand. The same instance when provisioned with a \gls*{ml} capacity block reserved in advance drops down to \$31.46/hour. % To illustrate the pros and cons of this approach, we can make an analogy with eating food at a restaurant.
% The current model for using cloud is akin to going to a restaurant with a buffet that charges by the plate, time spent eating, and by weight, and then allowing the consumer to enter and have dinner. The average consumer will likely come in to eat, and have their fill with only a general awareness of the number of plates they have consumed. The awareness might extend to the amount of food put on each plate, but it might not. Behind the scenes, there is a calculation that happens that looks at the exact weight of each different ingredient, and then presents the diner with a bill the next day. The diner might be surprised to realize they ate more of an expensive item than they intended, or dined for a longer time than was remembered. 
While the cost can be estimated via a pricing interface, it is difficult to account for the unpredictable nature of experiments. Prices also change across time and location. % Cost calculators may not fully capture the full set of smaller components that need to be put together that will assemble into the entire bill. 
% Often the experiment presents with surprises or simple variation in application running times. The simple act of 
% Further, running experiments on hugely expensive resources like GPU can add anxiety that hinders the user ability to efficiently execute them. 

%The consumer is hungry, and will likely do minimal planning or eyeballing and have most attention focused on eating the food or enjoying the experience. This same challenge exists with planning for using on-demand resources in cloud. 

Costs that vary based on configuration such as networking, virtual machine type, backups, region or zone, licenses, storage type, IOPs, and instance type are problematic because the consumer needs to be informed about product price variation. The additional investment in time translates into an additional monetary investment in data collection to better predict total costs. The barrier to entry to running cost-controlled experiments is too high. 
Using a real-time cost meter to track spending would improve transparency. % A real-time cost meter to track spending would be the equivalent. 
The delay in cost propagation can cause unnecessary uncertainty that does not provide awareness to unexpected spending.

\subsection{Price Comparison Tools}
While it may not be in a single cloud's best interest to compare their resources to other clouds, many tools have emerged (web interfaces and command line) that facilitate comparison. As an example of intra-cloud comparison, the \gls*{aws} region comparison tool \footnote{https://region-comparison.aws.com/} allows cost comparison of different resources. The SkyPilot project \footnote{https://github.com/skypilot-org/skypilot}  makes it easy and quick to compare GPU prices between clouds, and then to deploy \gls{ai}/\gls{ml}  workloads. A larger effort supported by cloud vendors to add comparison transparency would benefit the community of scientific users.

\subsection{Dark Patterns}

A dark pattern is a deliberate interface design that steers a user toward a vendor-desired outcome. An example is a company making a button with a desired outcome more visually apparent. In cloud, dark patterns are often associated with spend. A well-known example is related to egress costs. Although EU regulations, including the Data Act and the EU Cloud and AI Development Act \cite{UnknownUnknown-ey}, have phased out exit fees that could cause vendor lock-in, the cost of moving data remains significant, with estimates of adding 3.5-80x additional markup for egress bandwidth \footnote{https://blog.cloudflare.com/aws-egregious-egress/}.

These dark patterns are often related to awareness about billing. Although clouds have APIs that support cost estimation, they come with dark patterns. The Google Cloud billing API does not offer what the majority of users need, a cost per instance family per hour, but instead breaks down each family into units of memory (RAM), core hours, licenses and operating systems, and network. The developer user is required to map descriptions of families to instance names and sizes, and understand how to assemble different units of items in nanos \cite{google-cloud-money} to determine a monetary amount. A true desire to deliver transparency would not make this task so challenging. Web interfaces that are available can help, but are not programmatically parse-able. \gls*{aws} is more transparent, offering a price API that directly provides hourly prices for recognizable instance types.

While clouds offer services for budget alerts, many of these helper tools require enabling additional APIs and storage. The user is required to spend more money to understand how they are spending their money. This type of service places undue burden on the user to understand spending. Providing transparency should be the default. Cloud vendors likely do not have strong incentives to provide real-time transparency, either because greater awareness could lead to a reduction in spending, or because investing in the effort provides no technical benefit to them. 

The incentive for any cloud vendor to consider this model is trust. Lack of transparency can lead to a loss of trust for many new consumers. When cost awareness is hard to come by, especially in the context of purchasing complex services, a natural inclination is to assume a desire to mislead. Given that resources are invoiced down to the smallest unit, and that the cloud owns all of this information and can build services around exposing it, the responsibility to provide the information should arguably belong to cloud vendors.

\subsection{Spot Instances}

The spot instance model postulates that a cloud has spare capacity and allows its customer base to bid on it. Savings can be close to 90\% \cite{Unknown2025-gk}. Spot instances can be pre-empted at any time. The initial model of spot instance provisioning was based on bidding, however in 2017 \gls*{aws} changed their auction-based model to a retail-based ``price smoothing'' model that made the price-setting algorithm more opaque. Under this model, the customer does not have transparency into why an instance is terminated, and it was observed this led to overall higher prices, with some exceptions \cite{George2019-og}. A spot instance model could be paired with backfilling \cite{Srinivasan2003-ez}, allowing for smaller jobs to be run on resources that are idle, and pre-empting jobs when capacity is needed elsewhere. 

This is an example where overall transparency was reduced, and it is reasonable to infer that \gls*{aws} chose the approach to increase profits. 
% The perception of the initial model, which offered some transparency, might persist so that ``spot instances'' still carry somewhat of an illusion of transparency. 
\gls{aws} customers can use an API to get current and historical prices. The API limits to 50 requests a day, restricting the ability to model the underlying algorithm. Google Cloud does not offer an API. It becomes the customer's choice to use this model over ``on-demand'', a decision between paying a higher, stable price for a greater guarantee of availability. In the experience of this author, spot is best suited for smaller, independent jobs that can finish quickly, where losing an instance does not have a significant impact on the success of a workload or ensemble. Orchestration tools should be flexible to select either depending on the tradeoff between cost and time to completion dictated by the user.

\subsection{Reserved Instances and Payment Plans}

% Mindset of that person

Reserved instances are akin to a volume discount. Reserved instances require a customer to understand instance per-unit costs for a region, and to agree to a 1 or 3 year contract \cite{Slingerland2023-ya}. As commitment to a specific instance type can be rigid, \gls{aws} expanded their capabilities to Savings Plans, which also offer commitment-based discounts, but across a broader range of services and types. For both vehicles, the customer can choose to pay immediately, monthly, or at the end. Cloud vendors offer reserved instances, which are best for workloads that need to run constantly and those with consistent resource requirements, at discounts of up to 72\%. A downside of this model is that a reserved instance contract cannot be canceled, but only resold on a marketplace \cite{UnknownUnknown-bp}. Microsoft Azure has similar models, some of which list the same 72\% discount, but reservations can be exchanged or refunded up to \$50,000 (policy subject to change) \cite{UnknownUnknown-al}. Google Cloud calls this model a ``Committed Use Discount'' (CUD) \cite{UnknownUnknown-gf}, and only allows for monthly billing with no cancellation policy. There is typically a ``breaking point'' \cite{Carlin2019-fv} that can help to decide between using on-demand and reserved instances. If demand for instances is large enough, bulk purchase discounts may no longer be lucrative for clouds. 

A customer should perform a thorough examination of the subtle differences and contractual fine print of the policies and pricing models. Devising an optimal usage strategy would likely require consumers to create tools to track changes to configuration details and policies over time. The availability of information on policies and configurations is not enough to facilitate effective usage of cloud given its complexity and rapid rate of change.

\subsection{Reserved Capacity}
A capacity reservation is a request for a number of instances of a specific type, allocated immediately or between 5 and 120 days (\gls*{aws}) \cite{UnknownUnknown-xo} in the future. The allowed instance types for 120 day advanced reservations are limited to a smaller set, and the request must meet a minimum size of 100 vCPU. This means that getting highly contended resources for a study via a future reservation is often a manual process that requires human intervention and interaction between the customer and support. The  time slot availability can be during off-hours, and obtaining a full reservation is frequently impossible~\cite{sochat2025usabilityevaluationcloudhpc}. While capacity reservations bill at an on-demand price regardless of resource utilization, they can be canceled prior to the end of the reservation. \gls*{aws} also offers the ability to share capacity reservations with other accounts. 

Google Cloud offers a similar model to make reservations, either within one project or shared across projects, with additional features such as compact placement or \emph{auto-delete} to ensure that billing does not continue after the reservation period ends. Cancellation or deletion is only allowed before the reservation lock time \cite{UnknownUnknown-eo}, which is typically 8 weeks before the future reservation begins. It is not clear under what conditions this would be useful to a customer. The documentation hints at the option to submit a manual support request to see if the reservation can be canceled after it is locked. Azure offers on-demand capacity reservations \cite{UnknownUnknown-gg} that can be canceled at any time, yet many features such as proximity placement groups are not available.

\vspace{-3mm}

\subsection{Future Scheduling}
In 2025, \gls*{aws} introduced Capacity Blocks for ML that allow reservations of resource blocks at a future date. The service allows registration up to 6 weeks in advance, but sets a lower limit for a registration time of 24 hours. For a cluster of 32 nodes with H100 GPUs, this would cost approximately \$26,000 for a single day, a cost that is unlikely to be affordable for a scientific user. An alternative is capacity blocks with availability in the next 24 hours resulting from early reservation completion, which leaves the resources open to claim. Under this unintended side effect of the Capacity Blocks for ML model, it is possible to get resources for a smaller slice of time at a quantity that might be affordable under a science budget. Filling in cloud utilization gaps is analogous to backfilling in \gls*{hpc} resource managers, where smaller jobs are allowed to fill availability gaps \cite{Srinivasan2003-ez}. % The distinction with spot instances is that there is no pre-emption, as the customer pays full price. 
This kind of model exposed more officially and extended would benefit the scientific community. It could be possible to have scientific queues where small jobs are allowed to be submitted with resource needs and a time limit, and then automatically provisioned  when resources are available. % While the inner scheduler of a cloud is not known, for \gls*{aws}, arguably if the process described above is manual, this simply would need automation added to support.

The Google Cloud Dynamic Workload Scheduler (DWS) has a \emph{calendar mode} that works with Compute Engine reservations to allow resource reservations in the future \cite{Lohmeyer2023-hk}. The request for GPU is limited up to a count of 16, and the minimum time requirement is 24 hours. The resource types that can be reserved are limited, and enabling the feature requires interaction with a sales team. The user needs to be able to make a time-bounded request to be certain of a future start time.  % , and likely more profitable for cloud vendors to better utilize resources.  

\subsection{Serverless and Consumption Based Pricing}
Serverless or consumption-based pricing is often paired with \emph{functions as a service} and similar offerings. In these models, there is an operation (e.g., running a cloud function) with a fixed rate and the customer is charged for the number or quantity of operations. For example, for Google Cloud Run functions, request times are rounded to the nearest 100ms, and the price varies depending on the workload requiring GPU or not. The equivalent in \gls*{aws} is Lambda \cite{UnknownUnknown-ns}, which charges by the millisecond and does not support GPUs. Similarly to Lambda, Microsoft Azure functions do not appear to support GPUs. Performing a workload-specific cost and performance analysis is necessary to determine if a serverless model is preferred to a \gls{vm} deployment.

\section{Possible Futures}
In this next section we discuss possible futures, including pricing models and features that would support the scientific community to access resources. In some cases, there would be a clear benefit or incentive for the cloud vendor, and in other cases, more thinking is warranted.

\subsection{Micro-Commitments}

The concept of a savings plan can be modified to allow for shorter commitment durations and finer granularity. For example, a scientific user needing to run scaled experiments might use on-demand resources for small testing and development, and then require a dedicated set of resources, but in the order of weeks, days, or even specific hours. A micro-commitment would be a savings plan with shorter commitment durations and finer granularity that could better fit the budget of smaller research groups. Such a model would create a dynamic, short-term capacity market, and combine the predictability of reservations with the agility of on-demand.

\subsection{Rental of Unused Capacity}
For entities purchasing GPUs for private data centers, it has become lucrative to resell GPU capacity on online marketplaces oriented for that\footnote{https://www.latent.space/p/gpu-bubble}. Notably, these resale markets do not obviously include major cloud vendors. It would benefit the larger community if cloud vendors had reservation models that also allowed resale of unused capacity, as the entity that made the initial reservation would get money back, and a smaller entity that could not afford a large reservation would get compute time. This rental market could extend to time sharing of individual GPUs, however this introduces more security issues and potential risks if the shares are not completely isolated between customers. If the rental model is challenging to fit, these same unused resources could be made available to a queue intended for scientific use. A further incentive for large centers to engage in this kind of model could be economic incentives such as tax reductions or publicity.

\subsection{Tax Incentivized Fractal Sharing}
Rental of unused capacity can be further improved by changing the party responsible for the rental agreement. Much discussion of  policy places the responsibility to allocate resources on cloud vendors, unduly burdening the vendors. As an alternative, the responsibility to share a reservation might fall on the top level tier of cloud customers. Under such a model, a large entity that is able to afford purchasing a large reservation of GPUs, still facing the problem of having unused capacity, would receive a tax relief based on the portion of resources shared for academic or research pursuits. \gls{aws} gets close to this with its Nonprofit Credit Program\footnote{https://aws.amazon.com/government-education/nonprofits/nonprofit-credit-program/}, however the scope is limited to nonprofits with 501c3 designation, and educational institutions are not eligible. The annual promised funding (\$5k) could cover hosting costs, but likely would not cover scientific computing. 

A tax-incentivized sharing model could be used by academic groups that could pool together for purchasing power, but perhaps could not invest on an individual level. Tools could be created to allow collective groups to control scheduling queues and resource sharing. The cloud vendor could potentially increase profits if the design opens the market to smaller groups, and the top-level customer could receive a tax benefit. Assuming that large entities do not use resources at maximum capacity, there would be an overall improvement in resource utilization. %This model is similar to tax reliefs that are granted in Singapore for housing or living in proximity to elderly relatives \cite{UnknownUnknown-gg}.

\subsection{Predictive Scheduling and Future Reservation}
Predictive scheduling is the idea that a workflow tool can anticipate the future needs of a workflow. If a cloud API supports requests for reserving resources for future use, such as needing a GPU in 30 minutes, then a queue could be created to accept the requests. % Perhaps this request has a small cost to the user, and 
A future reservation would be made, with window of acceptance time directly before the full reservation. During this time, the workflow tool would need to accept and use the reservation. If the window passes and the reservation is not accepted, it would not negatively impact the cloud because the resource would return to the on-demand market. This model would supplement on-demand, allowing users to get access to resources within a transparent time frame without waiting (and incurring costs) if the exact count is not available at the exact time they are requested. This model would require workflow tools to anticipate future needs, and in the case of error make a request for an on-demand resource instead. Workflows that can allow for temporal gaps in execution would be the first to test. Such a model would require sophisticated scheduling on part of the cloud vendors, a variant of algorithms that already exists. There is opportunity for collaboration between cloud vendors and workload manager developers to create cloud queue infrastructure that works similarity and integrates well into \gls{hpc}, possibly allowing more seamless movement between environments.

\subsection{Re-emergence of Spot Blocks}
Amazon introduced the concept of \emph{spot blocks} in 2015 \cite{Unknown2015-sd} for ``defined duration workloads.'' A customer could use the same spot instance bidding model to get a guaranteed block of instances for a defined period of time. This model was discontinued in 2021, and it can only be speculated why. It could be that spot blocks were too similar to on-demand (and replacing it), essentially providing the same machines at a lower price and hurting profit. This might be an example where there is benefit to the customer, but not enough benefit to the cloud vendor leads to its extinction.

\subsection{Time Transparency}
A model that allows for future job scheduling enables time transparency -- the possibility to schedule a future block of resources. Unlike an on-demand request that would timeout after a fixed duration and partially allocate resources that incur cost, time transparency would show an estimated future time for a job. The consumer would have more certainty about scheduling than exists now, allowing for better workflow planning. The cloud vendor would be able to release resources into the market if they are not used. Making the cloud easier and less frustrating to use could increase customer satisfaction and adoption. Time transparency might also allow flexible market-driven cost models for resource allocation, and present an opportunity to collect more nuanced data about workloads. This model is similar to how traditional HPC workload managers already work. % There is likely already an internal queue that exists for customers within cloud providers, but it is not exposed due to not wanting to reveal that different customers receive different priority levels. 

%\subsection{Garage Clusters}
%A workaround to the issues with cloud costs is simply not to use cloud, and small entities that are unable to get the resources they need are experimenting with this model \cite{CNCF-Cloud-Native-Computing-Foundation-2024-mp}. More specifically, a ``garage cluster'' is a proposed design that sets up a central control plane or orchestrator in a locally owned resource, and then patches in geographically distributed resources from markets \cite{Cheah2024-cs} that are more cost effective and available. This model would have challenges around security and trust, as a hodgepodge of distributed machines from unknown data centers are not guaranteed to be trusted or secure. This model would also not do well for applications that require low latency connectivity, which usually requires machines to be physically close together.

\subsection{Backward Bursting}
In case GPUs are not available in the cloud, the ability to burst to on-premises resources with GPUs would enable rapid workflow execution. In this setup, % given a Kubernetes control plane running on a cloud 
a cloud resource or desired services could intelligently interact with nodes running on-premises. As an example, a control plane running in cloud would trigger provisioning of a kubelet on-premises, and sent credentials to join to the larger cluster. The biggest barrier for centers to adopt this approach would be a tendency to not want to expose ports to the outside world, which would generally be required for inter-node communication. \gls*{aws} has experimented with this idea via their \emph{hybrid node} model \footnote{https://docs.aws.amazon.com/eks/latest/userguide/hybrid-nodes-overview.html}, however this approach leads to institutions paying for using their own resources. If data transfer is required, strategies would be needed to minimize data movement costs between centers and cloud.

\subsection{Real-time Pricing}
If spending was reported in real time, consumer trust in cloud billing could improve. The drawback to cloud providers of this model is that transparency could lead to immediate customer awareness of spending and consequent reduction. However, providing this level of transparency could give consumers more confidence to be able to use cloud, and could lead to higher utilization when fear is dispelled. Lost trust resulting from unexpected spending could be restored.

\section{Conclusion}
High performance computing is at an inflection point. The future capability to do efficient, modern computational science results from access to computational resources. While science constitutes a small part of the customer base, the eventual returns on investment for scientific computing are outsized. The current economic model that prioritizes business needs for solving immediate problems does not account for advances in science that require longer time frames. However, these longer time frames afford flexibility. If the cloud could provide stronger guarantees about when work can occur, scientists can accommodate these future resources. We must advocate for integrated cost models that satisfy market needs without leaving scientific discovery behind. It is our responsibility to summarize the current landscape, describe what we need, and make suggestions for new, possible futures. We encourage discussion and collaboration to work toward a future of profitable solutions and successful science.

\begin{acks}
We would like to thank Dan Reed ``HPC Dan'' for fruitful discussion on early ideas. This work was performed under the auspices of the U.S. Department of Energy by Lawrence Livermore National Laboratory under Contract DE-AC52-07NA27344 (LLNL-JRNL-2007462).
\end{acks}

%%
%% The next two lines define the bibliography style to be used, and
%% the bibliography file.
\bibliographystyle{ACM-Reference-Format}
\bibliography{software}

\end{document}